# Systematic study of the electronic structure and the magnetic properties of a few-nm-thick epitaxial $(Ni_{1-x}Co_x)Fe_2O_4$ ($x = 0 - 1$) layers grown on $Al_2O_3(111)/Si(111)$ using soft X-ray magnetic circular dichroism: effects of cation distribution

Yuki K. Wakabayashi,[1,*,†] Yosuke Nonaka,[2] Yukiharu Takeda,[3] Shoya Sakamoto,[2] Keisuke Ikeda,[2] Zhendong Chi,[2] Goro Shibata,[2] Arata Tanaka,[4] Yuji Saitoh,[3] Hiroshi Yamagami,[3,5] Masaaki Tanaka,[1,6] Atsushi Fujimori,[2] and Ryosho Nakane[1,7,‡]

[1]*Department of Electrical Engineering and Information Systems, The University of Tokyo, 7-3-1 Hongo, Bunkyo-ku, Tokyo 113-8656, Japan*
[2]*Department of Physics, The University of Tokyo, Bunkyo-ku, Tokyo 113-0033, Japan*
[3]*Materials Sciences Research Center, Japan Energy Atomic Agency, Sayo, Hyogo 679-5148, Japan*
[4]*Department of Quantum Matters, ADSM, Hiroshima University, Higashi-Hiroshima 739-8530, Japan*
[5]*Department of Physics, Kyoto Sangyo University, Motoyama, Kamigamo, Kita-Ku, Kyoto 603-8555, Japan*
[6]*Center for Spintronics Research Network, Graduate School of Engineering, The University of Tokyo, 7-3-1 Hongo, Bunkyo-ku, Tokyo 113-8656, Japan*
[7]*Institute for Innovation in International Engineering Education, The University of Tokyo, 7-3-1 Hongo, Bunkyo-ku, Tokyo 113-8656, Japan.*
[*]Current affiliation: *NTT Basic Research Laboratories, NTT Corporation, 3-1 Morinosato-Wakamiya, Atsugi, Kanagawa 243-0198, Japan*
† Corresponding author: wakabayashi.yuki@lab.ntt.co.jp
‡ Corresponding author: nakane@cryst.t.u-tokyo.ac.jp

Abstract

We study the electronic structure and the magnetic properties of epitaxial $(Ni_{1-x}Co_x)Fe_2O_4(111)$ layers ($x = 0 - 1$) with thicknesses $d = 1.7 - 5.2$ nm grown on $Al_2O_3(111)/Si(111)$ structures, to achieve a high value of inversion parameter $y$ which is the inverse-to-normal spinel-structure ratio, and hence to obtain good magnetic properties even when the thickness is thin enough for electron tunneling as a spin filter. We revealed the crystallographic (octahedral $O_h$ or tetrahedral $T_d$) sites and the valences of the Fe, Co, and Ni cations using experimental soft X-ray absorption spectroscopy and X-ray magnetic circular dichroism spectra and configuration-interaction cluster-model calculation. In all the $(Ni_{1-x}Co_x)Fe_2O_4$ layers with $d$ ~4 nm, all Ni cations occupy the $Ni^{2+}$ ($O_h$) site, whereas Co cations occupy the three different $Co^{2+}$ ($O_h$), $Co^{2+}$ ($T_d$), and Co$^{3+}$ ($O_h$) sites with constant occupancies. According to these features, the occupancy of the Fe$^{3+}$ ($O_h$) cations decreases and that of the Fe$^{3+}$ ($T_d$) cations increases with decreasing $x$. Consequently, we obtained a systematic increase of $y$ with decreasing $x$ and achieved the highest $y$ value of 0.91 for the NiFe$_2$O$_4$ layer with $d$ = 3.5 nm. From the $d$ dependences of $y$ and magnetization in the $d$ range of 1.7 – 5.2 nm, a magnetically-dead layer is present near the NiFe$_2$O$_4$/Al$_2$O$_3$ interface, but its influence on the magnetization was significantly suppressed compared with the case of CoFe$_2$O$_4$ layers reported previously [Y. K. Wakabayasi et al., Phys. Rev. B **96**, 104410 (2017)], due to the high site selectivity of the Ni cations. Since our epitaxial NiFe$_2$O$_4$ layer with $d$ = 3.5 nm has a high $y$ values (0.91) and a reasonably large magnetization (180 emu/cc), it is expected to exhibit a strong spin filter effect which can be used for efficient spin injection into Si.

## I. INTRODUCTION

Towards the realization of Si-based spintronic devices for practical use at 300K [1-4], one of the most important building blocks is a spin injector/extractor which inject/extract a highly spin-polarized electrons into/from a Si channel. For such spin injector/detector junctions, the spin filter effect through an inverse spinel ferrite barrier is expected to be very useful, since it completely selects down-spin polarization of tunneling electrons by the spin-dependent tunnel probability which originates from the spin-polarized lower down-spin and higher up-spin conduction bands of the spinel ferrites. In addition, the Curie temperatures $T_C$ of bulk ferrites are typically far above room temperature: $T_C$ = 793 K for $CoFe_2O_4$ [5-7] and $T_C$ = 850 K for $NiFe_2O_4$ [8,9]. However, the experimental spin-polarization values estimated from the tunnel magnetoresistance ratio have been much less than the expectations in multi-layered structures with metal electrodes and a few-nm-thick ferrite tunnel barrier [8,10,11]. These results indicate that the ferrite barriers in these experiments had neither the ideally spin-polarized band structure nor good magnetic properties of the inverse spinel ferrites, owing to imperfection in cation ordering and lattice structures.

The inverse and normal spinel structures of $MFe_2O_4$ ($M$ = Co or Ni) ferrites are defined by the cation occupancies on the octahedral ($O_h$) and tetrahedral ($T_d$) sites in the lattice of O anions. Figure 1(a) shows a schematic picture of the spinel structure, where small red, small blue, and large gray spheres represent the $O_h$ sites, $T_d$ sites, and oxygen anions, respectively, and blue and red arrows represent the antiferromagnetic coupling between the magnetic moments of cations at the $T_d$ and $O_h$ sites, respectively. Generally, the inverse (normal) spinel structure stands for full occupancy of the $T_d$ sites by Fe ($M$) cations. To quantify the regularity of cation distribution, the inversion parameter $y$ is frequently defined by the ideal chemical formula $[M_{1-y}Fe_y]_{Td}[Fe_{2-y}M_y]_{Oh}O_4$ and represents the inverse-to-normal spinel-structure ratio: $y = 1$ ($y = 0$) denotes the perfect inverse (normal) spinel structure. Hereafter, the ratio of the total $O_h$ to the total $T_d$ sites occupied by Fe and $M$ cations is referred to as the total $O_h/T_d$ site ratio and is 2 in the above ideal chemical formula. From first-principles calculations for $MFe_2O_4$ ferrites with $y = 1$ [7,12], the lower down-spin conduction band is composed of $3d$ ($t_{2g}$) states of Fe cations at the $O_h$ sites, whereas the higher up-spin conduction band is composed of the $3d$ ($e$) states of Fe cations at the $T_d$ sites, as schematically shown in Fig. 1(b). When $y < 1$, up-spin midgap states are formed, which results in the decrease of spin selectivity during the electron tunneling [7,12,13] and also in the degradation of magnetic properties [6,13-17]. Therefore, the spin filter effect requires a thin $MFe_2O_4$ layer ($M$ = Co or Ni) with a high $y$. In addition to the above-mentioned cation distribution determined by the site selectivity of cations, a study on an epitaxial $Fe_3O_4$ layer on a MgO substrate [18] reported that the polar interface with oxide materials causes unoccupied $T_d$ sites, which should lead to a significant degradation of magnetic properties; The amount of unoccupied $T_d$ sites is maximum at the heterointerface and exponentially decreases with increasing the $Fe_3O_4$ thickness in the thickness range below 10 nm. Recently, we also confirmed such characteristics in epitaxial $CoFe_2O_4$(111) layers with various thicknesses $d$ = 1.4, 2.3, 4.0, and 11 nm on $Al_2O_3$(111)/Si(111) [19,20]; We found that a magnetically-dead layer originating from various complex networks of superexchange interactions is present mostly at the $CoFe_2O_4/Al_2O_3$ interface at $d$ = 1.4 nm, $y$ increases with increasing $d$, and the magnetization comparable to that of bulk materials were obtained only at $d$ = 11 nm. This is a serious problem in utilizing the spin filter effect, since the spin filter tunnel barrier requires a few-nm-thick ferrite with a high $y$ and good magnetic properties, and this is probably the reason why the spin polarization of electrons is smaller even at low temperatures (4 – 10 K) reported in Refs. [8,13].

The purpose of this study is to achieve a high $y$ and good ferrimagnetic properties even in a few-nm-thick ferrite layer epitaxially grown on $Al_2O_3$(111)/Si(111) by overcoming the above-mentioned problems. Our main material is $NiFe_2O_4$, since Ni cations are known to have a significantly higher site selectivity for the $O_h$ sites than Co cations: $y$ > 0.8 for bulk $NiFe_2O_4$ [21-23] and $y$ = 0.68–0.80 for bulk $CoFe_2O_4$ [24-27]. In the present work, in order to systematically reveal the effect of Ni cations on $y$ as well as the magnetic properties, we investigate the properties of $NiFe_2O_4$ layers with various thicknesses $d$ (= 1.7, 3.5, and 5.2 nm), compare with our previous report on $CoFe_2O_4$ [20], and also characterize an epitaxial NiO(111) layer to confirm the spectra of $Ni^{2+}$ cations occupying the $O_h$ sites. Furthermore, we also investigate the properties of $Ni_{1-x}Co_xFe_2O_4$ with $x$ = 0.25, 0.5, and 0.75, since such a study allows us to directly reveal the difference in the site selectivity between Co and Ni cations, and the $x$-dependent change in the magnetic properties and electronic structure will be useful information for optimizing the spin filter effect in Si-based tunnel junctions. Note that detailed properties, particularly the electronic structure, of $Ni_{1-x}Co_xFe_2O_4$ have never been clarified yet [28,29].

Our experimental techniques are soft X-ray absorption spectroscopy (XAS) and X-ray magnetic circular dichroism (XMCD), which are extremely sensitive tools to the local electronic structure and the magnetic properties of each element in magnetic materials [30-34], and allow us to determine the crystallographic sites and valences of cations [35-37]. In addition, since XMCD is free from the diamagnetic signal from the substrate, one can perform accurate measurements on ultra-thin magnetic layers. Therefore, XMCD measurements are useful for the systematic investigation of a few-nm-thick magnetic layers in oxide magnetic multilayers.

In this paper, we present the electronic structure and magnetic properties of epitaxial $NiFe_2O_4$(111) layers ($d$ = 1.7, 3.5, and 5.2 nm) and epitaxial 3.5-nm-thick $(Ni_{1-x}Co_x)Fe_2O_4$(111) layers ($x$ = 0.25, 0.5, 0.75) grown on $Al_2O_3$(111)/Si(111) structures using XAS and XMCD. The crystallographic sites and valences of the cations in the layers are determined using the experimental XAS and XMCD spectra and theoretical calculation based on the configuration-interaction (CI) cluster model [38]. We obtained high $y$ values of 0.79 – 0.91 for $NiFe_2O_4$ layers with $d$ = 1.7, 3.5, and 5.2 nm, owing to the 100% selectivity of $Ni^{2+}$ for the $O_h$ sites. We found that the high site selectivity of Ni cations and the low site selectivity of Co cations are universal nature in all the $(Ni_{1-x}Co_x)Fe_2O_4$ layers, which leads to the $x$-dependent occupancies of the $Fe^{3+}$ cations for the $T_d$ and $O_h$ sites. Consequently, $y$ systematically increases with decreasing $x$ and it shows the highest value at $x$ = 0 ($NiFe_2O_4$).

**II. EXPERIMENTAL**

We grew two series of epitaxial thin films: (i) Epitaxial single-crystalline $NiFe_2O_4$(111) layers with various thicknesses $d$ (= 1.7, 3.5, and 5.2 nm), and (ii) 3.5-nm-thick $(Ni_{1-x}Co_x)Fe_2O_4$(111) layers with various $x$ (= 0.25, 0.5, and 0.75), on 1.4-nm-thick $\gamma$-$Al_2O_3$(111) buffer layer / $n^+$-Si(111) substrates using pulsed laser deposition (PLD), as shown in Fig. 1(c). For a reference, we grew a 3.5-nm-thick NiO layer on a 1.4-nm-thick $\gamma$-$Al_2O_3$(111) buffer layer / $n^+$-Si(111) substrate by PLD, to confirm the spectra of $Ni^{2+}$ ($O_h$) cations. During the growth of these layers, the substrate temperature (500°C), $O_2$ pressure (10 Pa), and the laser setup were the same as those in our previous report [20]. The stoichiometry of each layer is referred to as that of a sintered target used for each growth.

Figure 2 shows reflection high-energy electron diffraction (RHEED) patterns of an epitaxial $NiFe_2O_4/\gamma$-$Al_2O_3$(111)/Si(111) structure after the growth of a 5.2-nm-thick $NiFe_2O_4$ layer as an example. The RHEED patterns of all the $(Ni_{1-x}Co_x)Fe_2O_4$ ($x$ = 0, 0.25, and 0.75) layers show similar sharp 6-fold streaks with a 2×2 reconstruction. These patterns indicate a high-quality 2-dimensional epitaxial growth mode in all the $(Ni_{1-x}Co_x)Fe_2O_4$ layers. The RHEED patterns of the NiO layer also show 6-fold spotty streaks. These 6-fold symmetry diffraction patterns indicate that one domain is completely aligned with the Si substrate with the epitaxial relationship of $(Ni_{1-x}Co_x)Fe_2O_4[11\bar{2}](111)$ // $\gamma$-$Al_2O_3[11\bar{2}](111)$ // $Si[11\bar{2}](111)$ and $NiO[11\bar{2}](111)]$ // $\gamma$-$Al_2O_3[11\bar{2}](111)$ // $Si[11$ 2̄](111), whereas another domain is rotated by 60° in the (111) plane. These double domain structures are basically the same as those reported in Refs. [19,20].

Figures 3(a) and (b) show cross-sectional high-resolution transmission electron microscopy (HRTEM) images of the NiFe$_2$O$_4$ film with $d$ = 5.2 nm projected along the Si <11$\bar{2}$> axis. Almost the entire region of the NiFe$_2$O$_4$ layer has an epitaxially-grown single-crystalline structure with a smooth surface and interface with the γ-Al$_2$O$_3$ buffer layer. A ~2-nm-thick SiO$_x$ interfacial layer seems to be formed by the high O$_2$ pressure of 10 Pa during the growth of the NiFe$_2$O$_4$ layer. The orange dashed lines represent anti-phase boundaries (APBs), which are growth defects of the cation sublattice inherent in the spinel structure [14-16]: The oxygen lattice remains unchanged across an APB whereas the cation sublattice is shifted by the <220> translation vector [16].

We performed XAS and XMCD measurements at the soft X-ray beamline BL23SU of SPring-8 with a twin-helical undulator of in-vacuum type [39], which allows us to perform efficient and accurate measurements of XMCD with various incident photon energies. The monochromator resolution was $E/\Delta E$ > 10000. XMCD spectra were obtained by reversing the photon helicity at each energy point and were recorded in the total-electron-yield (TEY) mode. To eliminate possible experimental artifacts, we averaged XMCD spectra taken for both positive and negative magnetic fields applied perpendicular to the layer surface. The direction of the incident X-rays was also perpendicular to the layer surface. Backgrounds of the XAS spectra at the Fe, Co, and Ni $L_{2,3}$ edges were subtracted from the raw data, assuming that they are hyperbolic tangent functions. All the measurements were performed at 300 K, and a magnetic field applied perpendicular to the layer surface is denoted by $\mu_0 H$. Note that the data of the CoFe$_2$O$_4$ layer [(Ni$_{1-x}$Co$_x$)Fe$_2$O$_4$ with $x$ = 1] are the same as those reported previously [20].

To clarify the correlation between the cation distribution and magnetic properties quantitatively, we determined the crystallographic sites and valences of Fe, Co, and Ni cations using the experimental XAS and XMCD spectra and cluster-model calculation. It has been well recognized that the XAS and XMCD spectra of transition-metal oxides strongly depend on the 3$d$ electron configuration, crystal field, spin-orbit coupling, and electron-electron interaction within the transition-metal cation, and the hybridization of 3$d$ electrons with other valence electrons. Taking into account these effects, we calculated XAS and XMCD spectra for Fe, Co, and Ni cations with a specific site and valence by employing the CI cluster model [38]. In the calculation, we adopted empirical relationship between the on-site Coulomb energy $U_{dd}$ and the 3$d$-2$p$ hole Coulomb energy $U_{dc}$: $U_{dc} / U_{dd}$ = 1.25 [40], and that between the Slater-Koster parameters $pd\sigma$ and $pd\pi$: $pd\sigma / pd\pi$ = -2.17 [41]. The hybridization strength between O 2p orbitals $T_{pp}$ was fixed to be 0.7 eV (for $O_h$ site) and 0 eV (for $T_d$ site) [38,40], and 80% of the ionic Hartree-Fock values were used for Slater integrals. Thus, the crystal-field splitting $10Dq$, the charge-transfer energy $\Delta$, $U_{dd}$, and $pd\sigma$ were treated as adjustable parameters. As for the Fe cations, $10Dq$, $\Delta$, and $pd\sigma$ were adjusted to reproduce the various experimental Fe $L_{2,3}$-edge spectra by the weighted sum of calculated spectra for the $Fe^{3+}$ ($O_h$), $Fe^{3+}$ ($T_d$), and $Fe^{2+}$ ($O_h$) cations. $U_{dd}$ was fixed to the value reported for $Fe_3O_4$ [40]. As for the Ni cations, all the adjustable parameters were chosen to reproduce the experimental spectra of $NiFe_2O_4$ layers measured with $\mu_0H$ = 7 T. As for the Co cations, the adjustable parameters were adopted from previously reported $CoFe_2O_4$ values [20]. The parameter values used for these calculations are listed in Table 1. The spin magnetic moment $m_{spin}$ and the orbital magnetic moment $m_{orb}$ were also calculated within the CI cluster model using the above parameters for the Fe, Co, and Ni cations, and they are summarized in Table 2.

**III. EXPERIMENTAL RESULTS AND ANALYSES**

**A. Thickness $d$ dependence of the cation distribution and magnetic properties in the $NiFe_2O_4$ layers**

Figure 4 (a) and (b) show Fe and Ni $L_{2,3}$-edge XMCD ($\mu^+ - \mu^-$) spectra normalized at 708.7 and 851.1 eV, respectively, for the $NiFe_2O_4$ layers with $d$ = 1.7, 3.5, and 5.2 nm measured with $\mu_0H$ = 7 T. In Fig. 4(b), a normalized Ni $L_{2,3}$-edge XMCD spectrum for the NiO layer with $\mu_0H$ = 10 T is also shown as a reference (black curve). Here, $\mu^+$ and $\mu^-$ denote the absorption coefficients for the photon helicities parallel and antiparallel to the Ni 3d majority spin direction, respectively. These spectra show multiplet structures which are characteristic of the localized 3d state of Fe and Ni cations in oxides [9,35-37]. The Fe $L_3$-edge XMCD spectra have a positive peak at 708.0 eV and two negative peaks at 706.8 and 708.7 eV. It is well known that those peaks at 706.8, 708.0, and 708.7 eV mainly come from $Fe^{2+}$ ($O_h$), $Fe^{3+}$ ($T_d$), and $Fe^{3+}$ ($O_h$) cations, respectively [20,35,40], where the superscript number denotes the valence of the cations. Our calculation also supports these assignments, as will be described. The peak height for the $Fe^{3+}$ ($T_d$) cations (708.0 eV) are comparable to or larger than that for the $Fe^{3+}$ ($O_h$) cations (708.7 eV) in all the $NiFe_2O_4$ layers, indicating that the amount of the $Fe^{3+}$ ($T_d$) cations is comparable to or larger than that of the $Fe^{3+}$ ($O_h$) cations. This result means that all the $NiFe_2O_4$ layers have a high inversion parameter $y$. On the other hand, the XMCD signals at the Ni $L_3$ edge are mostly negative. This means that the spin magnetic moments of the $Fe^{3+}$ ($T_d$) cations and Ni cations have an antiparallel configuration, as shown in Fig. 1(a), which is characteristic of the Ni ($O_h$) cations in inverse spinel ferrites [20,42,43].

In Fig. 4(b), the Ni $L_{2,3}$-edge XMCD spectra for various $d$ are identical with each other and are similar to that for the NiO layer that has only $Ni^{2+}$ ($O_h$) cations. This indicates that all the $NiFe_2O_4$ layers have $Ni^{2+}$ ($O_h$) cations. Meanwhile, the normalized XMCD intensities for the $Fe^{3+}$ ($T_d$) cations (708.0 eV) in the $d$ = 3.5 and 5.2 nm layers are the same, and decreases with decreasing $d$ from 3.5 to 1.7 nm, as shown in Fig. 4(a), indicating that the amount of the $Fe^{3+}$ ($T_d$) cations relative to the total amount of all the Fe cations becomes smaller for $d$ = 1.7 nm. These results mean that the total $O_h/T_d$ site ratio is larger than 2 near the $NiFe_2O_4/Al_2O_3$ interface. This feature was also observed at the $Fe_3O_4/MgO$ interface [18].

We quantitatively estimated the crystallographic sites and the valences of the Fe and Ni cations in the $NiFe_2O_4$ layers using the experimental XAS [$(\mu^+ + \mu^-)/2$] and XMCD spectra and the CI cluster-model calculations. Figures 5(a) and (b) show calculated Fe $L_{2,3}$-edge XAS and XMCD spectra for the $Fe^{3+}$ ($O_h$), $Fe^{3+}$ ($T_d$), and $Fe^{2+}$ ($O_h$) cations, respectively, using the parameters in Table 1. Figures 5(c) and (d) show the experimental Fe $L_{2,3}$-edge XAS and XMCD spectra measured with $\mu_0H$ = 7 T for the $NiFe_2O_4$ layers with $d$ = 1.7, 3.5 and 5.2 nm, and the corresponding curve fitting (particularly for the $L_3$ edges) with the weighted sum of the calculated spectra shown in Figs. 5(a) and (b). Here, the weight of the each calculated spectra represents the site occupancy of each Fe cations. The experimental spectra are well reproduced by the weighted sum of the calculated spectra, including the characteristic kink structure at around 707.5 eV in the XMCD spectra. These results give strong evidence that the Fe cations are composed of the $Fe^{3+}$ ($O_h$), $Fe^{3+}$ ($T_d$), and $Fe^{2+}$ ($O_h$) cations. From these fits, we obtained the magnetic moments of the $Fe^{3+}$ ($O_h$), $Fe^{3+}$ ($T_d$), and $Fe^{2+}$ ($O_h$) cations with $\mu_0H$ = 7 T and the site occupancy of each Fe cations. We also estimated the inversion parameter $y$ from the site occupancies of these Fe cations; $y = 2f[Fe^{3+}$ ($T_d$)], where $f[Fe^{3+}$ ($T_d$)] represents the site occupancy of the $Fe^{3+}$ ($T_d$) cations [9]. We note that the above estimation of $y$ is based on the ideal chemical formula $[M_{1-y}Fe_y]_{Td}[Fe_{2-y}M_y]_{Oh}O_4$; the effects of oxygen vacancies, non-stoichiometry, and/or the change of the total $O_h/T_d$ site ratio are ignored for the estimation of $y$ values.

In the same way, the experimental Ni $L_{2,3}$-edge XAS and XMCD spectra for the $NiFe_2O_4$ layers with $d$ = 1.7, 3.5, and 5.2 nm are also well reproduced by the calculated spectra for the $Ni^{2+}$ ($O_h$) cation using the parameters in Table 1, as shown in Figs. 6(a)–(d). This result confirms that all the $NiFe_2O_4$ layers have only $Ni^{2+}$ ($O_h$) cations. From these fits, we also obtained the magnetic moment of the $Ni^{2+}$ ($O_h$) cations with

$\mu_0 H = 7$ T.

Figures 7(a) shows the $d$ dependence of the site occupancies of $Fe^{3+}$ ($O_h$), $Fe^{3+}$ ($T_d$), and $Fe^{2+}$ ($O_h$) cations in the $NiFe_2O_4$ layers, which were estimated from the same fitting procedure. In the same figure, the inversion parameter $y$ estimated form the site occupancy of these Fe cations, as described above, is also shown. Considering the charge neutrality, the valence of all the Fe cations for all $d$ was essentially 3+ due to the fact that the occupancy of the $Ni^{2+}$ ($O_h$) cations is 100% in all $d$, as shown in Fig. 7(b). However, the occupancy of the $Fe^{2+}$ ($O_h$) cations was found to be ~5% in all $d$, which may come from small amount of the oxygen vacancies and/or non-stoichiometry in the $NiFe_2O_4$ layer. Whereas the occupancies of all the cations are almost the same for $d = 3.5$ and 5.2 nm, the occupancy of the $Fe^{3+}$ ($O_h$) cations becomes larger and that of the $Fe^{3+}$ ($T_d$) cations becomes smaller when $d$ decreases from 3.5 to 1.7 nm. In response to the site occupancies and site ratio, $y$ is as high as ~0.91 for $d = 3.5$ and 5.2 nm, but it becomes 0.79 when $d$ decreases from 3.5 to 1.7 nm.

Figure 8 shows XMCD intensity – $H$ curves measured at the Fe $L_3$ edge for the $NiFe_2O_4$ layers with $d = 1.7$ and 3.5 nm, in which each XMCD intensity along the vertical axis is scaled so that it represents the sum of the estimated magnetizations of all the Fe and Ni cations in each layer. The rhombus in the figure represents the magnetization at $\mu_0 H = 7$ T for the $NiFe_2O_4$ layer with $d = 5.2$ nm, which is scaled in the same manner. The magnetization at $\mu_0 H = 7$ T is correlated with $y$ in Fig. 7(a): it is ~180 emu/cc for $d = 3.5$ and 5.2 nm, and it decreases to 60 emu/cc when $d$ decreases from 3.5 to 1.7 nm.

**B. Co concentration $x$ dependence of the cation distribution and magnetic properties in the $(Ni_{1-x}Co_x)Fe_2O_4$ layers with $d = 3.5$ or 4 nm**

Figures 9(a), (b), and (c) show Fe, Co, and Ni $L_{2,3}$-edge XMCD spectra normalized at 708.7, 777.5, and 851.1 eV, respectively, for the 3.5 nm-thick $(Ni_{1-x}Co_x)Fe_2O_4$ layers ($x = 0, 0.25, 0.5,$ and 0.75) and the 4 nm-thick $CoFe_2O_4$ layer ($x = 1$) measured with $\mu_0 H = 7$ T. The XMCD signals at the Co and Ni $L_3$ edges are mostly negative, indicating that the magnetic moments of the Co and Ni cations have an antiparallel configuration to those of the $Fe^{3+}$ ($T_d$) cations, as shown in Fig. 1(a). This is characteristic of the Co ($O_h$) and Ni ($O_h$) cations in inverse spinel ferrites [20,35,42,43]. In Fig. 9(a), the normalized Fe $L_3$-edge XMCD intensity for the $Fe^{3+}$ ($T_d$) cations (708.0 eV) decreases with increasing $x$, namely, $y$ decreases with increasing $x$. In contrast, the normalized Co and Ni XMCD spectra do not vary with $x$, as shown in Figs. 9(b) and (c), respectively, which indicates that the site occupancies of the Ni and Co cations are constant for various $x$. Thus, the higher $y$ value with lower $x$ is associated with the difference of the site selectivity of the Ni and Co cations.

To clarify the correlation between the site selectivity, cation distribution, and magnetization quantitatively, we also determined the crystallographic sites and valences of Fe, Co, and Ni cations in the $(Ni_{1-x}Co_x)Fe_2O_4$ layers using the experimental XAS and XMCD spectra and the CI cluster-model calculation. Figures 10(a) and (b) show the experimental Fe $L_{2,3}$-edge XAS and XMCD spectra measured with $\mu_0H = 7$ T for the 3.5-nm-thick $(Ni_{1-x}Co_x)Fe_2O_4$ layers ($x = 0, 0.25, 0.5$, and $0.75$) and the 4-nm-thick $CoFe_2O_4$ layer ($x = 1$), and the corresponding curve fittings (particularly for the $L_3$ edges) with the weighted sum of the calculated spectra in Figs. 5(a) and (b). The experimental spectra are well reproduced by the weighted sum of the calculated spectra, confirming that the $Fe^{3+}$ ($O_h$), $Fe^{3+}$ ($T_d$), and $Fe^{2+}$ ($O_h$) cations are present. From these fits, we can obtain the site occupancies of the $Fe^{3+}$ ($O_h$), $Fe^{3+}$ ($T_d$), and $Fe^{2+}$ ($O_h$) cations as shown in Fig.12(a). We also estimated the inversion parameter $y$ from the site occupancies of these Fe cations.

Figures 11(a) and (b) show the calculated Co $L_{2,3}$-edge XAS and XMCD spectra for the $Co^{2+}$ ($O_h$), $Co^{2+}$ ($T_d$), and $Co^{3+}$ ($O_h$) cations, respectively, using the parameters in Table 1. Figures 11(c) and (d) show the experimental Co $L_{2,3}$-edge XAS and XMCD spectra measured with $\mu_0H = 7$ T for the 3.5 nm-thick $(Ni_{0.25}Co_{0.75})Fe_2O_4$ layer, and the weighted sum of the calculated spectra in Figs. 11(a) and (b). The experimental XAS and XMCD spectra are well reproduced by the weighted sum of the calculated spectra, which indicates that most of the Co cations are $Co^{2+}$ ($O_h$), $Co^{2+}$ ($T_d$), and $Co^{3+}$ ($O_h$) cations. The small discrepancy between the experimental spectra and the weighted sum of the calculated spectra may come from factors other than the assumptions in the calculation [20]: some other Co cations, such as low spin Co cations [44], Co cations at the trigonal prism sites [45], and Co cations under local distortion. The experimental Co $L_{2,3}$-edge XAS and XMCD spectra for the 3.5 or 4 nm-thick $(Ni_{1-x}Co_x)Fe_2O_4$ layers with other $x$'s ($x = 0.25, 0.5$, and 1) can also be reproduced by the same weighted sum of the calculated spectra since they are identical with those for the $(Ni_{0.25}Co_{0.75})Fe_2O_4$ layer in Fig. 9(b). From these calculations, we obtain the site occupancies of the $Co^{2+}$ ($O_h$), $Co^{2+}$ ($T_d$), and $Co^{3+}$ ($O_h$) cations at $\mu_0H = 7$ T as shown in Fig.12(b).

In the same way, the experimental Ni $L_{2,3}$-edge XAS and XMCD spectra for the 3.5 nm-thick $(Ni_{1-x}Co_x)Fe_2O_4$ layers with $x$ (= 0.25, 0.5, and 0.75), which are identical with that for the $NiFe_2O_4$ layer [Fig. 9(c)], also can be reproduced by the calculated spectra for the $Ni^{2+}$ ($O_h$) cation in Figs. 6(a) and (b). From these fits, we can also obtain the site occupancies of the $Ni^{2+}$ ($O_h$) cations at $\mu_0H = 7$ T as shown in Fig.12(c).

In Fig. 12(a), the inversion parameter $y$ estimated from the site occupancies of the Fe cations, as described above, is also shown by red rhombuses. In all the layers, all Ni cations occupy the Ni$^{2+}$ ($O_h$) site [Fig. 12(c)], whereas Co cations occupy the three different Co$^{2+}$ ($O_h$), Co$^{2+}$ ($T_d$), and Co$^{3+}$ ($O_h$) sites with constant occupancies [Fig. 12(b)]. On the other hand, the site selectivity of the Fe cations strongly depends on $x$: as $x$ decreases from 1, the occupancy of the Fe$^{3+}$ ($O_h$) cations decreases and that of the Fe$^{3+}$ ($T_d$) cations increases [Fig. 12(a)]. Although these changes are only ~10% in the full $x$ range, $y$ drastically increases with decreasing $x$ and it shows the highest value of 0.91 at $x = 0$.

**IV. DISCUSSION**

In this section, we summarize the experimental results described in the previous sections, and discuss the relationship between the electronic structure and the magnetic properties of the (Ni$_{1-x}$Co$_x$)Fe$_2$O$_4$ layers with $x = 0 - 1$, in order to provide a comprehensive understanding.

In all the (Ni$_{1-x}$Co$_x$)Fe$_2$O$_4$ layers with $d$ ~4 nm, all the Ni cations occupy the Ni$^{2+}$ ($O_h$) site, whereas Co cations occupy the three different Co$^{2+}$ ($O_h$), Co$^{2+}$ ($T_d$), and Co$^{3+}$ ($O_h$) sites with constant occupancies. This indicates that the coexistence of Ni and Co has no influence on the site selectivity of each cation, and that the amount of the Co$^{2+}$ ($T_d$) cations decreases and the amount of the Ni$^{2+}$ ($O_h$) cations increases as $x$ decreases. This means that the amount of the $M$ ($T_d$) cations decreases and the amount of the $M$ ($O_h$) cations increases with decreasing $x$. On the other hand, the site selectivity of the Fe cations strongly depends on $x$; as $x$ decreases from 1, the occupancy of the Fe$^{3+}$ ($O_h$) cations decreases and that of the Fe$^{3+}$ ($T_d$) cations increases, resulting in the highest $y$ value in the NiFe$_2$O$_4$ layer. This is simply understood by the ideal chemical formula $[M_{1-y}Fe_y]_{Td}[Fe_{2-y}M_y]_{Oh}O_4$; the increase in the amount of the Fe ($T_d$) cations originates from both the decrease in the amount of the $M$ ($T_d$) cations and the increase in the amount of the $M$ ($O_h$) cations with decreasing $x$.

In our previous report on the CoFe$_2$O$_4$ layers [20], we concluded that the degradation of magnetization in the thinner thickness ($d \leq 4$ nm) mainly originates from the magnetically-dead layer near the CoFe$_2$O$_4$/Al$_2$O$_3$ interface due to the decrease of $y$, reflecting the increase both in the site occupancy of the Co$^{2+}$($T_d$) cations and in the total $O_h/T_d$ site ratio from 2. In the case of the NiFe$_2$O$_4$ layers, the increase in the site occupancy of the Ni ($T_d$) cations is excluded due to the Ni$^{2+}$ ($O_h$) cations with 100% occupancy for any $d$. Although the $d$ dependences of $y$ (Fig. 7) and magnetization (Fig. 8) indicate the increase of the $O_h/T_d$ site ratio near the NiFe$_2$O$_4$/Al$_2$O$_3$ interface, the $y$ values for the NiFe$_2$O$_4$ layers are high 0.79 – 0.91. It should be noteworthy that $y = 0.79$ for $d = 1.7$ nm is slightly higher than $y = 0.75$ obtained for the CoFe$_2$O$_4$ layer with $d = 11$ nm, which is the highest value among the CoFe$_2$O$_4$ layers with $d = 1.4 - 11$ nm. Thus, the significant improvement in $y$ was achieved using NiFe$_2$O$_4$ even when $d$ is thin enough for electron tunneling. Furthermore, the improvement of the magnetic properties is confirmed by the magnetization at 7 T; the magnetization of the NiFe$_2$O$_4$ layers with $d = 3.5$ and 5.2 nm is 67% of that for a bulk material 270 emu/cc [46], whereas the magnetization of the 11-nm-thick CoFe$_2$O$_4$ layer was 44% of that for a bulk material 425 emu/cc [46].

On the other hand, despite such improvement, $y < 1$ and the magnetizations of the NiFe$_2$O$_4$ layers are still smaller than the magnetization of a bulk material, which indicates the presence of a magnetically-dead layer near the NiFe$_2$O$_4$/Al$_2$O$_3$ interface. In the NiFe$_2$O$_4$ layers, APBs are present, as shown in Fig. 3, and other structural defects may be also present, particularly in the vicinity of the domain boundary between the two in-plane-rotated domains. Thus, we concluded that the increase of the $O_h/T_d$ site ratio as well as other structural defects form the magnetically-dead layer moderately degrades the properties of the NiFe$_2$O$_4$ layers.

## V. CONCLUSION

We have investigated the electronic structure and the magnetic properties of the epitaxial NiFe$_2$O$_4$ layers with various thicknesses ($d = 1.7$, 3.5, and 5.2 nm) and (Ni$_{1-x}$Co$_x$)Fe$_2$O$_4$ layers with $x$ (= 0.25, 0.5, 0.75, and 1) and $d = 3.5$ or 4 nm in the epitaxial (Ni$_{1-x}$Co$_x$)Fe$_2$O$_4$(111)/Al$_2$O$_3$(111)/Si(111) structures using XAS and XMCD. We have also determined the crystallographic sites and the valences of the Fe, Co, and Ni cations using the experimental XAS and XMCD spectra, and the CI cluster-model calculation. In all the (Ni$_{1-x}$Co$_x$)Fe$_2$O$_4$ layers with $d \sim 4$ nm, all the Ni cations occupy the Ni$^{2+}$ ($O_h$) site, whereas Co cations occupy the three different Co$^{2+}$ ($O_h$), Co$^{2+}$ ($T_d$), and Co$^{3+}$ ($O_h$) sites with constant occupancies. According to these facts, the occupancy of the Fe$^{3+}$ ($O_h$) cations decreases and that of the Fe$^{3+}$ ($T_d$) cations increases with decreasing $x$. These features result in the highest $y$ value (= 0.91) in the NiFe$_2$O$_4$ layer. For the NiFe$_2$O$_4$ layers with $d = 1.7$, 3.5, and 5.2 nm, we obtained high $y$ values of 0.79 – 0.91, which are higher than the highest $y = 0.75$ obtained for the 11-nm-thick CoFe$_2$O$_4$ layer in our previous report. From the $d$ dependence of $y$ and magnetization, the magnetically-dead layer near the NiFe$_2$O$_4$/Al$_2$O$_3$ interface is also present in the NiFe$_2$O$_4$ layers, but it was found to have less influence on the magnetization compared with the case of CoFe$_2$O$_4$. Consequently, the improvements of $y$ and magnetization were achieved even in thinner $d$, mostly owing to the 100% $Ni^{2+}$ selectivity for $O_h$ sites. We also discussed that the increase of the $O_h/T_d$ site ratio together with other structural defects, such as APBs, are possible origins of the slight degradation of $y$ and the moderate degradation of magnetization.

As stated in the introduction, the spin filter effect requires a thin $M$Fe$_2$O$_4$ layer ($M$ = Co or Ni) with a high $y$. Judging from this standard, our NiFe$_2$O$_4$ layers are expected to exhibit a strong spin filter effect even when the thickness is thin enough for electron tunneling ($d$ = 3.5 nm). The results revealed in this study give us an important guideline for the design of Si-based spintronic devices using spinel ferrites; Among (Ni$_{1-x}$Co$_x$)Fe$_2$O$_4$ with $x$ = 0 – 1, NiFe$_2$O$_4$ is most promising to realize a highly-efficient spin injection into Si by the spin filter effect through a thin-layer spinel ferrite. For further improvement, interface engineering is necessary, such as introducing the epitaxial strain [42], to realize the ideal $O_h/T_d$ site ratio near the NiFe$_2$O$_4$/Al$_2$O$_3$ interface.

**ACKNOWLEDGEMENTS**

This work was partly supported by Grants-in-Aid for Scientific Research (Grants Nos. 26289086, 15H02109, and 17H02915), including the Project for Developing Innovation Systems from Ministry of Education, Culture, Sports, Science and Technology (MEXT), the CREST Program of JST (No. JPMJCR1777), Tohoku University, and the Spintronics Research Network of Japan (Spin-RNJ). This work was performed under the Shared Use Program of Japan Atomic Energy Agency (JAEA) Facilities (Proposal No. 2016B-E20) supported by JAEA Advanced Characterization Nanotechnology Platform as a program of "Nanotechnology Platform" of MEXT (Proposal No. A-16-AE-0030). The synchrotron radiation experiments were performed at the JAEA beamline BL23SU in SPring-8 (Proposal No. 2016B3841). Y. K. W. acknowledges financial support from Japan Society for the Promotion of Science (JSPS) through the Program for Leading Graduate Schools (MERIT) and the JSPS Research Fellowship Program for Young Scientists. S. S. acknowledges financial support from JSPS through the Program for Leading Graduate Schools (ALPS). A.F. is an adjunct member of Center for Spintronics Research Network.

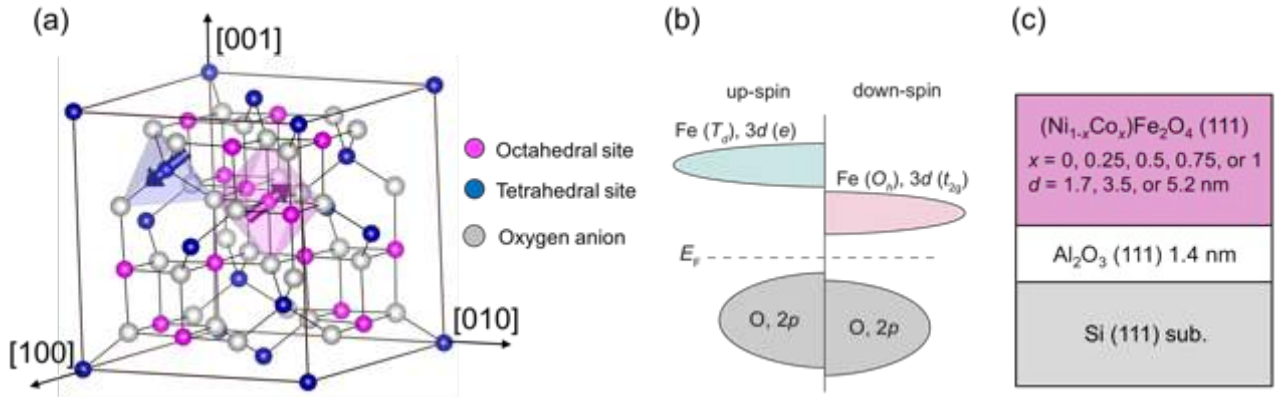

FIG. 1. Schematic pictures of (a) the spinel structure with the octahedral ($O_h$) and tetrahedral ($T_d$) sites, (b) the density of states of the valence-band top and conduction-band bottom for $M$Fe$_2$O$_4$ ($M$ = Co or Ni) ferrites, and (c) the sample structure. (a) Small red, small blue, and large gray spheres represent the $O_h$ sites, $T_d$ sites, and oxygen anions, respectively, and blue and red arrows represent the antiferromagnetic coupling between the magnetic moments of cations at the $T_d$ and $O_h$ sites. (b) The contributions of other orbitals are omitted for the simplicity.

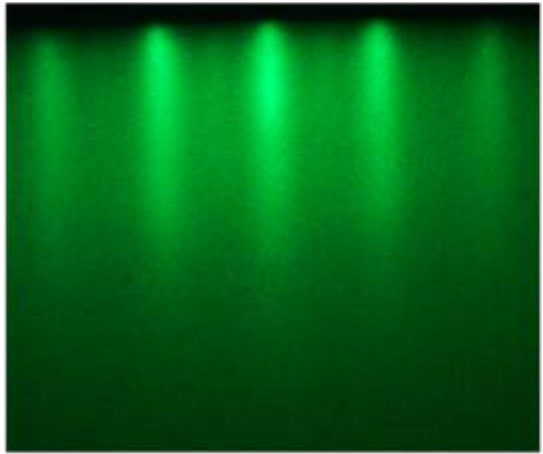 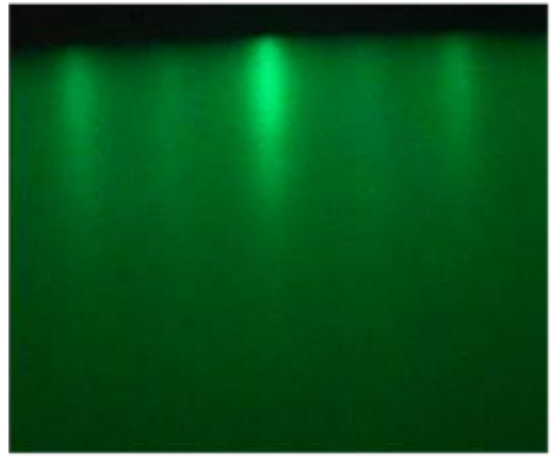

FIG. 2. Reflection high-energy electron diffraction (RHEED) patterns of an epitaxial NiFe$_2$O$_4$/γ-Al$_2$O$_3$(111)/Si(111) film after the growth of a 5.2-nm-thick NiFe$_2$O$_4$ layer by pulsed laser deposition (PLD), where electrons are incident along the (a) [1$\bar{1}$0] and (b) [11$\bar{2}$] directions of the Si substrate, respectively.

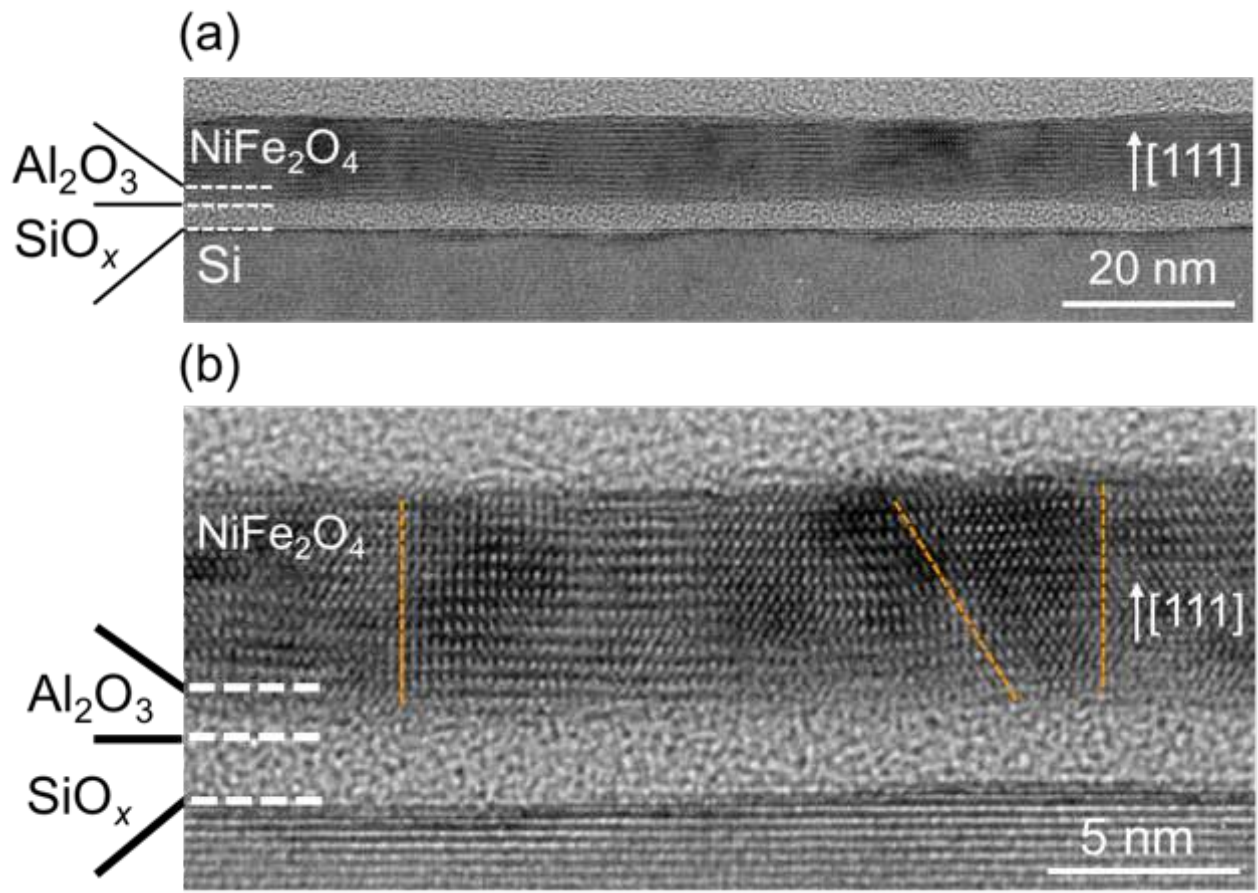

FIG. 3. High-resolution transmission electron microscopy (HRTEM) lattice image of the $NiFe_2O_4$ film with $d$ = 5.2 nm projected along the Si $<11\bar{2}>$ axis. (b) Magnified image of (a), where the orange dashed lines represent anti-phase boundaries (APBs).

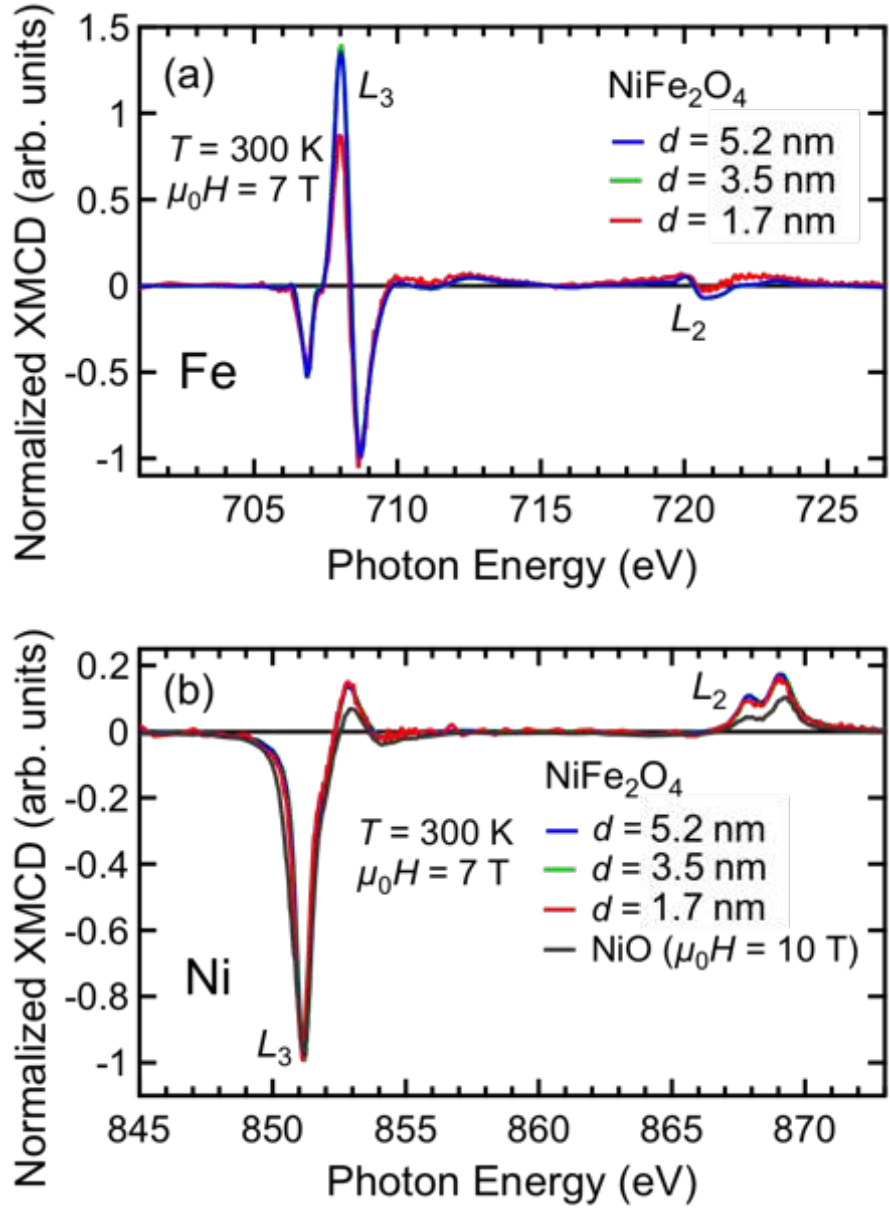

FIG. 4. Fe (a) and Ni (b) $L_{2,3}$-edge XMCD (= $\mu^+ - \mu^-$) spectra normalized at 708.7 and 851.1 eV, respectively, for the NiFe$_2$O$_4$ layers with $d$ = 1.7, 3.5, and 5.2 nm measured with a magnetic field $\mu_0 H$ = 7 T. In (b), a spectrum for a NiO layer measured with a magnetic field $\mu_0 H$ = 10 T is also shown.

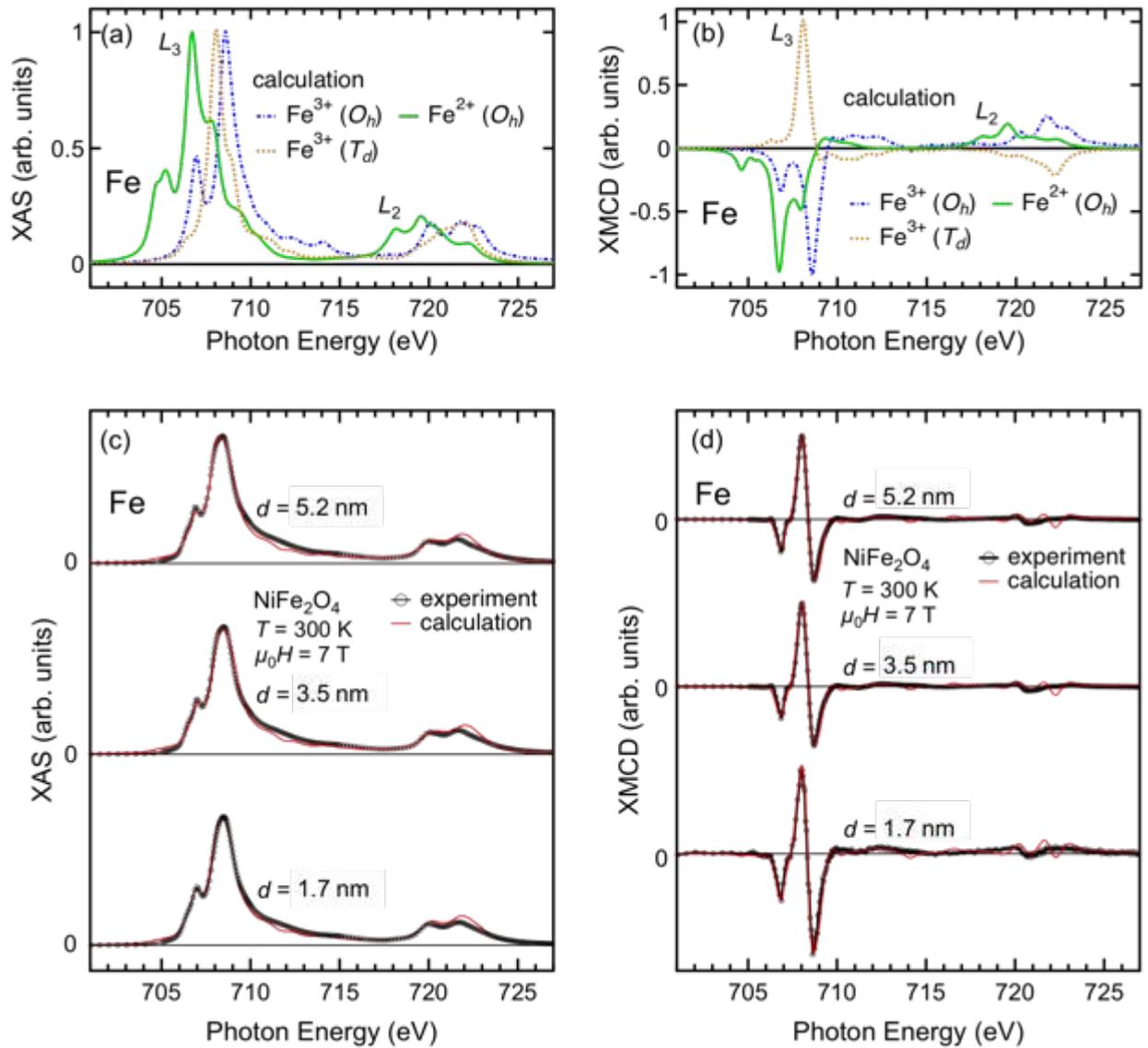

FIG. 5. Calculated Fe $L_{2,3}$-edge XAS [$(\mu^+ + \mu^-)/2$] (a) and XMCD (b) spectra, where the dot-dashed, dotted, and solid curves represent the spectra for $Fe^{3+}$ ($O_h$), $Fe^{3+}$ ($T_d$), and $Fe^{2+}$ ($O_h$), respectively. Experimental Fe $L_{2,3}$-edge XAS (c) and XMCD (d) spectra for the $NiFe_2O_4$ layers with $d$ = 1.7, 3.5, and 5.2 nm measured with a magnetic field $\mu_0 H$ = 7 T. In the figure, circles are experimental data and red curves are the weighted sum of the calculated spectra shown in panels (a) and (b). Each spectrum has been arbitrarily scaled for easy comparison.

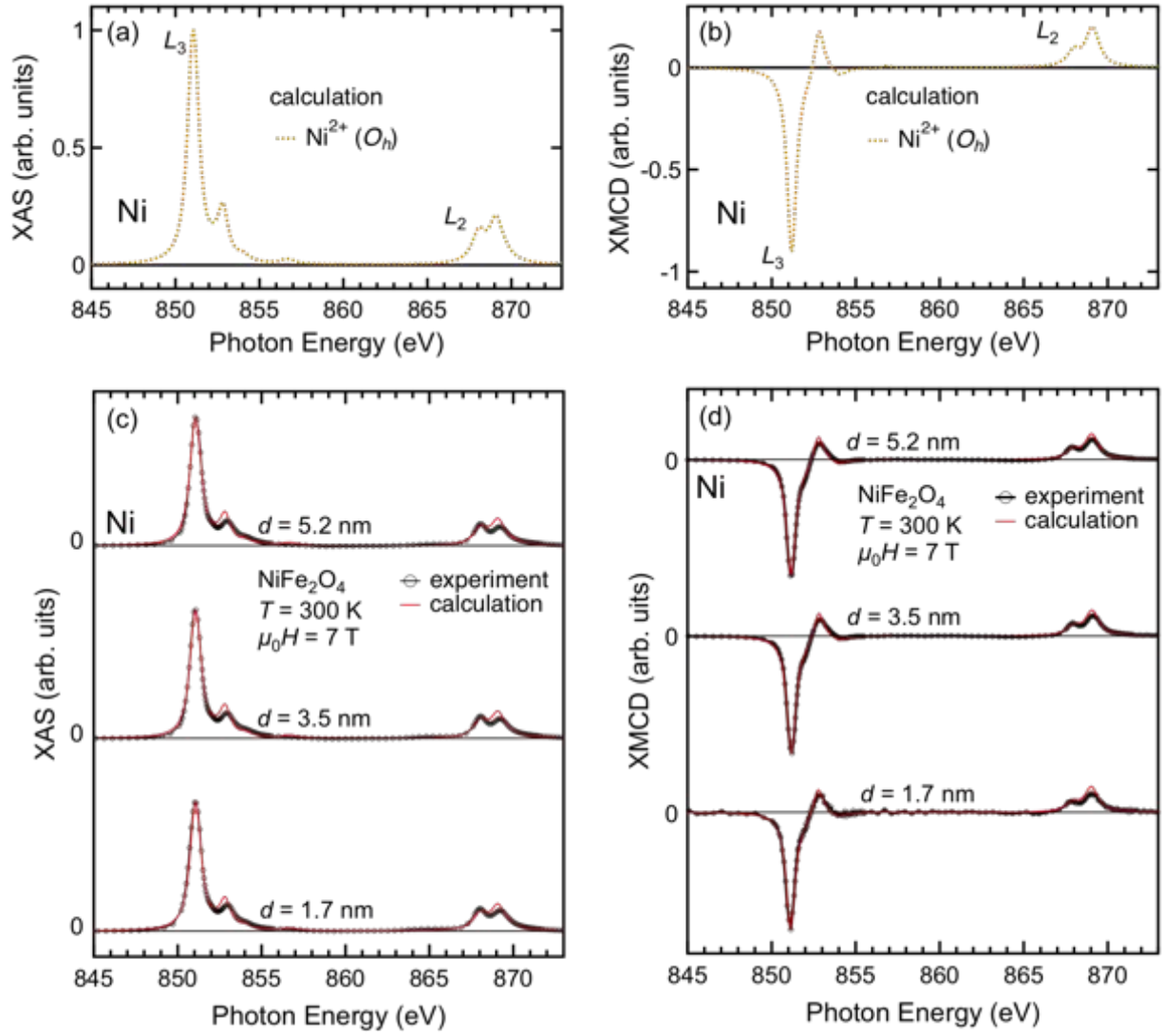

FIG. 6. Calculated Ni $L_{2,3}$-edge XAS (a) and XMCD (b) spectra for $Ni^{2+}$ ($O_h$). Experimental Fe $L_{2,3}$-edge XAS (c) and XMCD (d) spectra for the $NiFe_2O_4$ layers with $d$ = 1.7, 3.5, and 5.2 nm measured with a magnetic field $\mu_0H = 7$ T. In the figure, circles are experimental data and red curves are the weighted sum of the calculated spectra shown in panels (a) and (b). Each spectrum has been arbitrarily scaled for easy comparison.

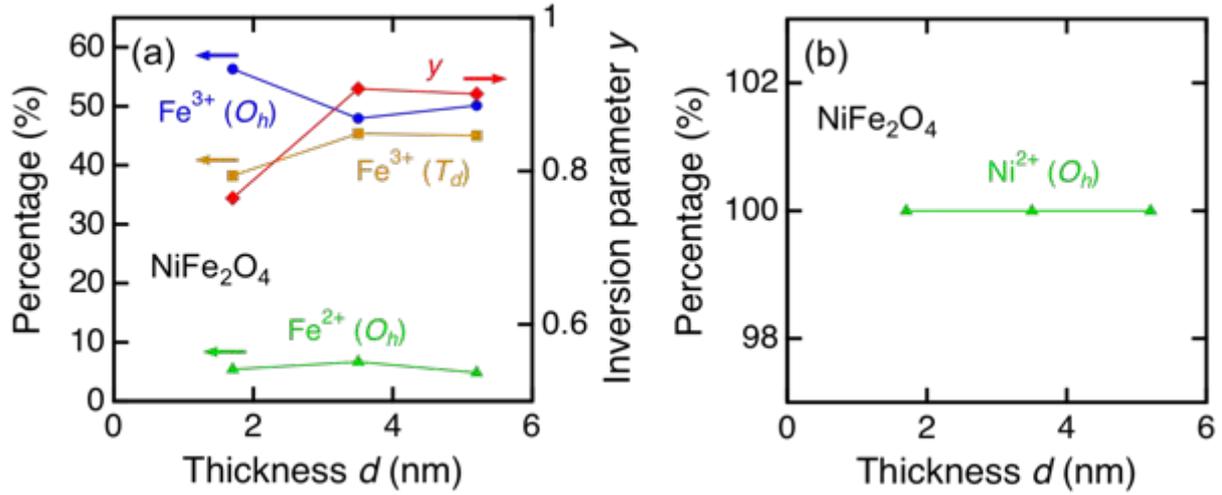

FIG. 7. (a) Thickness $d$ dependence of site occupancies of the Fe cations, where the circles, squares, and triangles represent $Fe^{3+}$ ($O_h$), $Fe^{3+}$ ($T_d$), and $Fe^{2+}$ ($O_h$), respectively, in the $NiFe_2O_4$ layers. Inversion parameter $y$ is also shown by rhombuses. (b) Thickness $d$ dependence of site occupancy of the Ni cations in the $NiFe_2O_4$ layers.

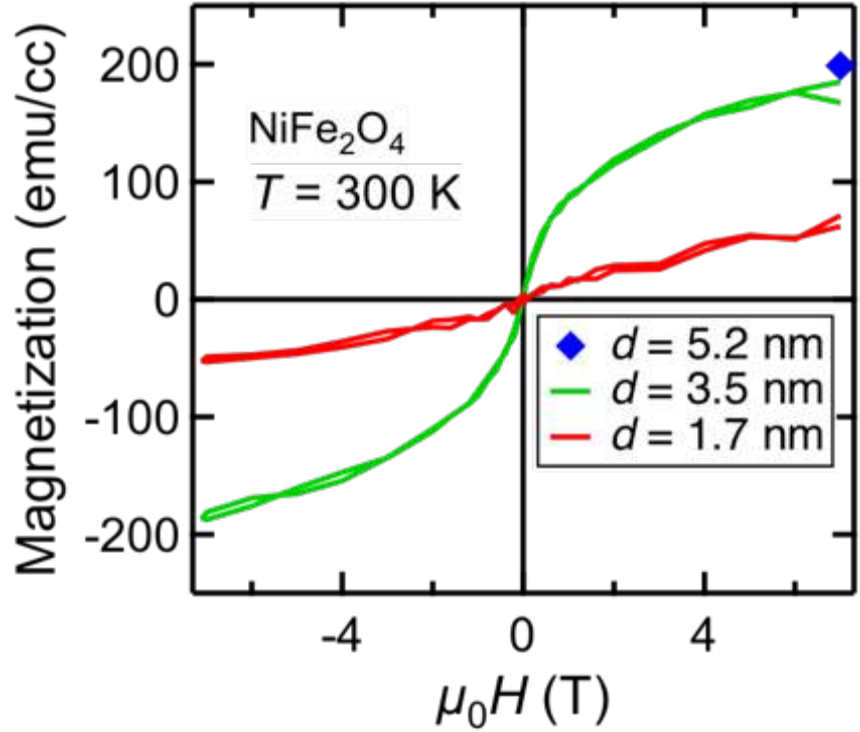

FIG. 8. XMCD intensity – $H$ curves measured at the Fe $L_3$ edge for the NiFe$_2$O$_4$ layers with $d = 1.7$ and 3.5 nm. The magnetization for the NiFe$_2$O$_4$ layers with $d = 5.2$ nm with $\mu_0 H = 7$ T is also shown by a rhombus. The vertical axis of the XMCD intensities is scaled so that it represents the sum of the magnetizations of the Fe and Ni cations estimated from the fits in Figs. 5(c), 5(d), 6(c), and 6(d).

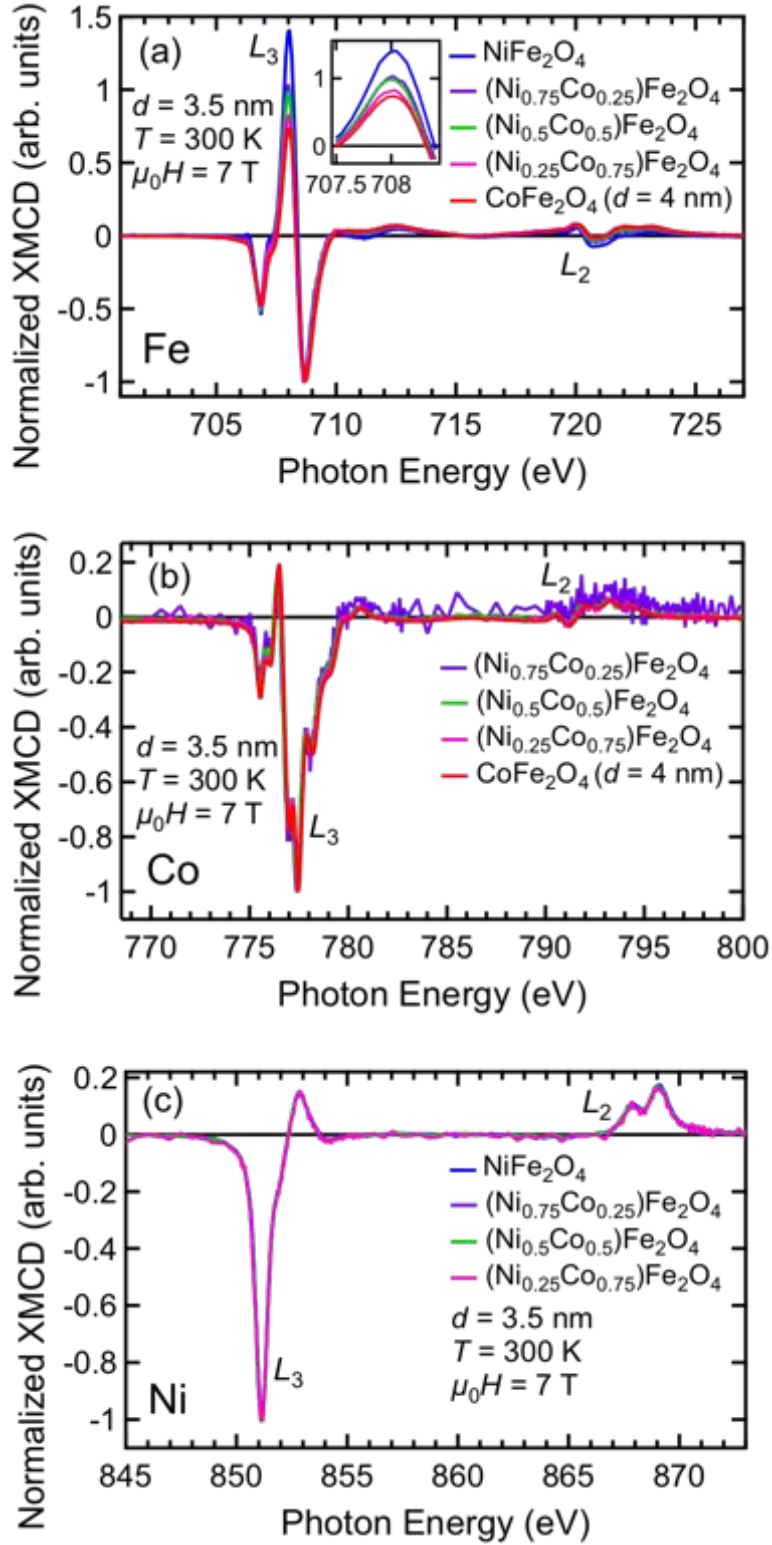

FIG. 9. Fe (a), Co (b), and Ni (c) $L_{2,3}$-edge XMCD spectra normalized at 708.7, 777.5, and 851.1 eV, respectively, for the $(Ni_{1-x}Co_x)Fe_2O_4$ layers with $x$ (= 0, 0.25, 0.5, 0.75, and 1) measured with a magnetic field $\mu_0H = 7$ T, where $d = 3.5$ nm for $x = 0 - 0.75$ and $d = 4.0$ nm for $x = 1$ ($CoFe_2O_4$). The inset of (a) shows magnified plots of the spectra at the Fe $L_3$ edges.

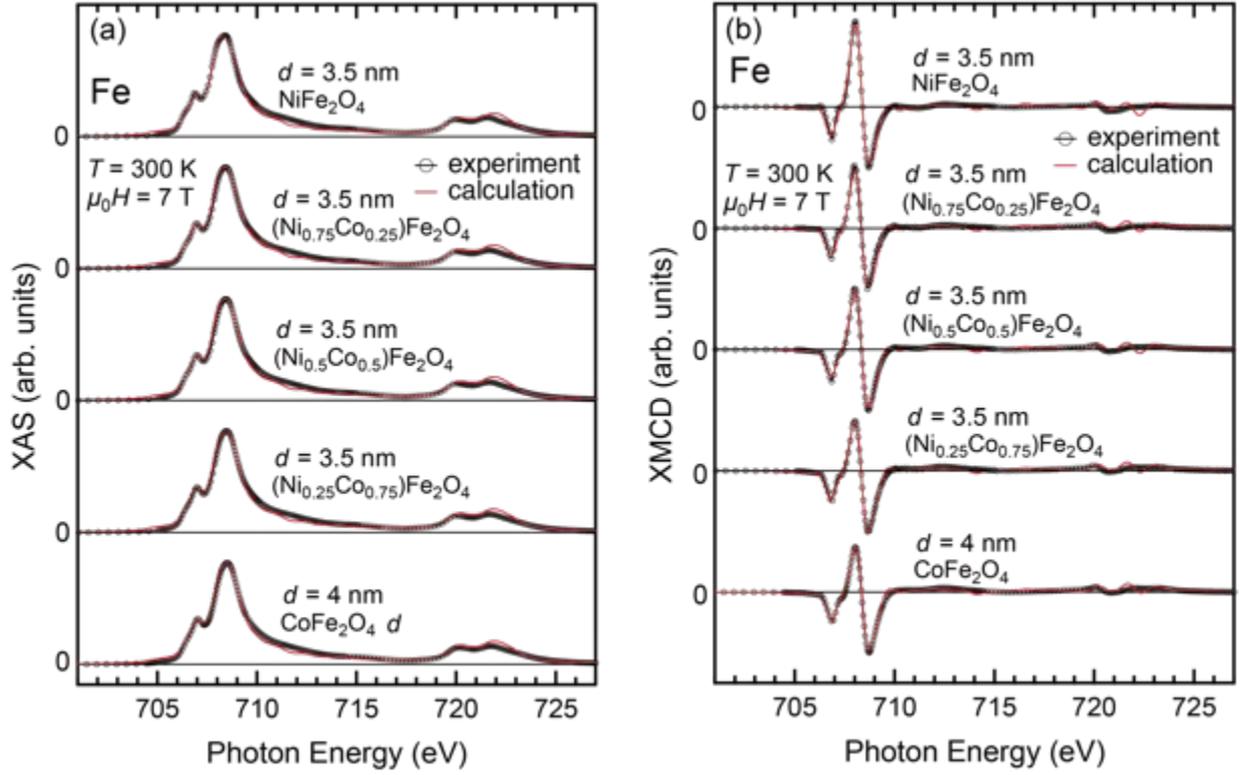

FIG. 10. Experimental Fe $L_{2,3}$-edge XAS (a) and XMCD (b) spectra for the $(Ni_{1-x}Co_x)Fe_2O_4$ layers (the same samples in Fig. 9) measured with a magnetic field $\mu_0H = 7$ T. In the figure, circles are experimental data and red curves are the weighted sum of the calculated spectra shown in Figs. 5(a) and 5(b). Each spectrum has been arbitrarily scaled for easy comparison.

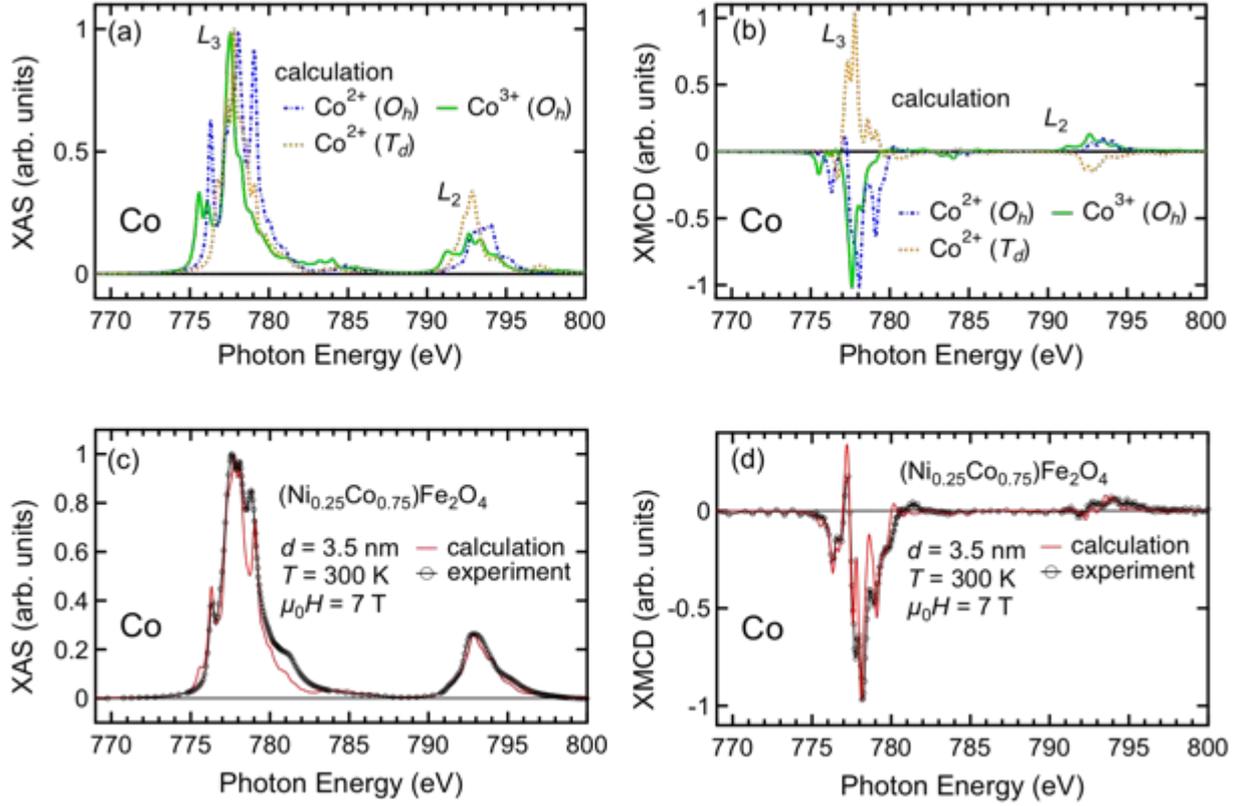

FIG. 11. Calculated Co $L_{2,3}$-edge XAS (a) and XMCD (b) spectra, where the dot-dashed, dotted, and solid curves represent the spectra for $Co^{2+}$ ($O_h$), $Co^{2+}$ ($T_d$), and $Co^{3+}$ ($O_h$), respectively. Experimental Co $L_{2,3}$-edge XAS (c) and XMCD (d) spectra for the $(Ni_{1-x}Co_x)Fe_2O_4$ layers (the same samples in Fig. 9) measured with a magnetic field $\mu_0 H = 7$ T. In the figure, circles are the experimental data and red curves are the weighted sum of the calculated spectra shown in panels (a) and (b).

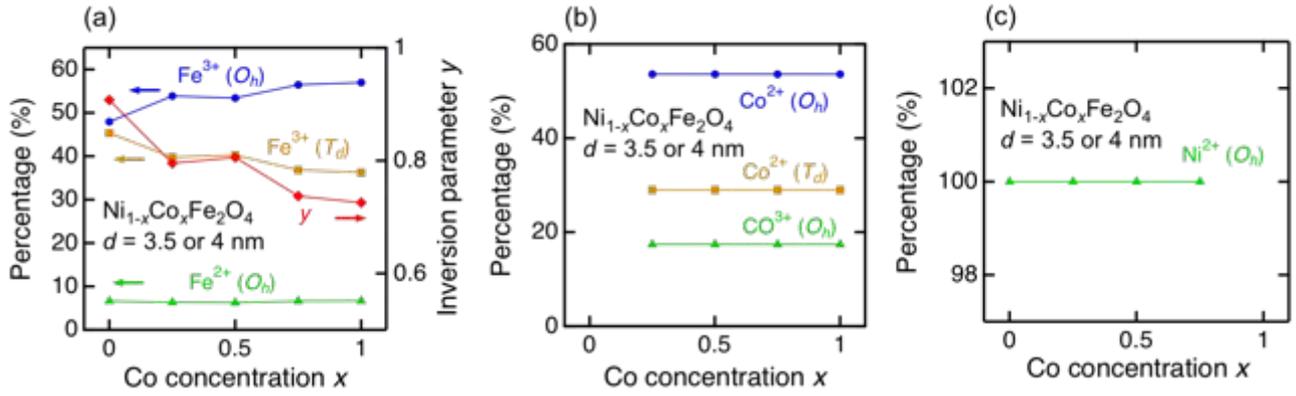

FIG. 12. Co concentration $x$ dependence of site occupancies of the Fe, Co, and Ni cations in the $(Ni_{1-x}Co_x)Fe_2O_4$ layers (the same samples in Fig. 9): (a) Fe cations, (b) Co cations, and (c) Ni cations. (a) Circles, squares, and triangles represent the site occupancies of the $Fe^{3+}$ ($O_h$), $Fe^{3+}$ ($T_d$), and $Fe^{2+}$ ($O_h$) cations, respectively, and the rhombuses denote inversion parameter $y$. (b) Circles, squares, and triangles represent the site occupancies of the $Co^{2+}$ ($O_h$), $Co^{2+}$ ($T_d$), and $Co^{3+}$ ($O_h$) cations, respectively. (c) Triangles represent the site occupancy of the $Ni^{2+}$ ($O_h$) cations.

|  | Δ | 10Dq | pdσ | $U_{dd}$ |
|---|---|---|---|---|
| $Fe^{3+}$ ($O_h$) | 0.4 | 0.9 | 1.2 | 6.0 |
| $Fe^{3+}$ ($T_d$) | 2.5 | -0.1 | 2.2 | 6.0 |
| $Fe^{2+}$ ($O_h$) | 6.5 | 0.9 | 1.6 | 6.0 |
| $Co^{2+}$ ($O_h$) | 5.6 | 0.5 | 1.3 | 6.5 |
| $Co^{2+}$ ($T_d$) | 6 | -0.3 | 1.4 | 6.0 |
| $Co^{3+}$ ($O_h$) | 0 | 0.5 | 1.3 | 6.0 |
| $Ni^{2+}$ ($O_h$) | 4.2 | 1.0 | 1.0 | 6.9 |

Table 1. Parameter values in units of eV used in the calculation based on the CI cluster model. For the Fe cations, $U_{dd}$ was adopted from [40].

|  | $Fe^{3+}$ ($O_h$) | $Fe^{3+}$ ($T_d$) | $Fe^{2+}$ ($O_h$) | $Co^{2+}$ ($O_h$) | $Co^{2+}$ ($T_d$) | $Co^{3+}$ ($O_h$) | $Ni^{2+}$ ($O_h$) |
|---|---|---|---|---|---|---|---|
| $m_{spin}$ ($\mu_B$/atom) | 4.41 | 4.42 | 3.62 | 2.54 | 2.87 | 3.20 | 1.67 |
| $m_{orb}$ ($\mu_B$/atom) | 0.02 | 0.01 | 0.56 | 1.07 | 0.47 | 0.70 | 0.26 |

Table 2. $m_{spin}$ and $m_{orb}$ ($\mu_B$/atom) calculated based on the CI cluster model for the $Fe^{3+}$ ($O_h$), $Fe^{3+}$ ($T_d$), $Fe^{2+}$ ($O_h$), $Co^{2+}$ ($O_h$), $Co^{2+}$ ($T_d$), $Co^{3+}$ ($O_h$), and $Ni^{2+}$ ($O_h$) cations.